\documentclass{ptephy}

\usepackage{delarray}
\usepackage{multirow}
\usepackage{amsmath,amssymb}
\usepackage{wrapfig}
\usepackage{color}
\usepackage{subfig}
\usepackage{dcolumn}  
\usepackage{lscape}
\usepackage{setspace}
\usepackage{textcomp}
\usepackage{feynmp}
\usepackage{here}

\allowdisplaybreaks[3]

\newcommand{\mydif}{\mathrm{d}}

\begin{document}
\title{The anomalous four-point interaction in the radiative leptonic $\tau$ decay}

\newcommand{\todai}{Department of Physics, University of Tokyo, Tokyo 113-0033}
\newcommand{\binp}{Budker Institute of Nuclear Physics SB RAS, Novosibirsk 630090}

\author{\name{N.~Shimizu}{1}, \name{D.~Epifanov}{2,3}, \name{J.~Sasaki}{1}}
\address{
\affil{1}{Department of Physics, University of Tokyo, Tokyo 113-0033}
\affil{2}{Budker Institute of Nuclear Physics SB RAS, Novosibirsk 630090}
\affil{3}{Novosibirsk State University, Novosibirsk 630090}
}

\begin{abstract}
As one of the extensions of the Standard Model, 
we investigate the anomalous four-point $\tau -W-\nu_\tau -\gamma$  
scalar- and tensor-type interactions, which originate from the 
gauge invariant dimension-five operators. 
The coupling constants are constrained by the measured 
branching ratio of the  $\tau^- \rightarrow \mu^- \nu_\tau \bar{\nu}_\mu \gamma$ decay: 
$-4.9 < \kappa_{\tau}^S <9.4$ and $-1.4 < \kappa_{\tau}^T<2.8$ at the 95\% confidence 
level for the scalar and tensor interactions, respectively. 
\end{abstract}
\subjectindex{B40, B50} 
\maketitle

\section{Introduction}

The decays of $\tau$ lepton provide unique opportunities to search for
the effects beyond the Standard Model (BSM) \cite{tau1,tau2,tau3,tau4,tau5,tau6,tau7,tau8,tau9,tau10,tau11}. The large mass of $\tau$ 
($m_{\tau}=(1776.86\pm 0.12)$~MeV/$c^2$ \cite{PDG_paper}), in comparison with that 
of the electron or muon, allows one to expect an essential enhancement 
in the sensitivity to the effects of New Physics (NP) \cite{eidel}. 

Of all tau decays, the leptonic ones are precisely calculated within 
electroweak sector of the SM, hence they offer a clean laboratory 
to search for the effects of NP. 
   
Through the measurement of Michel parameters, $\rho$, $\eta$, $\xi$ and $\xi\delta$, in  
ordinary leptonic decays $\tau^-\to\ell^-\nu_\tau\bar{\nu}_\ell$ ($\ell = e,~\mu$),
the experimental verification of the Lorentz structure of the charged weak interaction 
was carried out \cite{PDG_paper}. The most precision measurements were done by ALEPH \cite{etalep} 
and CLEO \cite{rhoCLEO} collaborations. 
 
The measurement of the ratio of the branching fractions 
$\mathcal{Q}\equiv\mathcal{B}(\tau^-\rightarrow\mu^-\nu_\tau\bar{\nu}_\mu)/\mathcal{B}(\tau^-\rightarrow e^-\nu_\tau\bar{\nu}_e)$ 
is used to test the lepton universality: 
\begin{align}
\left(\frac{c_\mu}{c_e}\right)^2=\frac{\mathcal{B}(\tau^- \rightarrow \mu^- \nu_\tau \bar{\nu}_\mu)}{\mathcal{B}
(\tau^- \rightarrow e^- \nu_\tau \bar{\nu}_e)}\frac{f(m_e^2/m_\tau^2)}{f(m_\mu^2/m_\tau^2)}, \label{eqratio}
\end{align}
where $c_\mu$ and $c_e$ are the couplings of $\tau$ with $\mu$ and $e$, respectively, 
$m_\ell$ ($\ell = e,~\mu$) is outgoing lepton mass, and $f(m_\ell^2/m_\tau^2)$ 
is a known function~\cite{citeratio}. The most precise measurement at $BABAR$, 
$\mathcal{Q}=0.9796\pm 0.0016 \pm 0.0036$~\cite{cite_babarratio}, is consistent 
with lepton universality. 
 
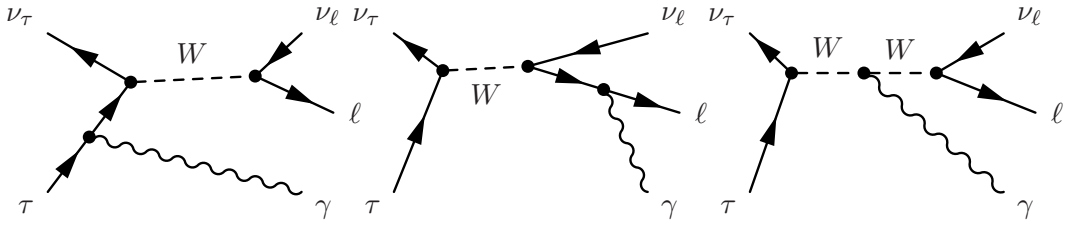
\begin{figure}[]
\begin{center}
\begin{fmffile}{externalph1}
\begin{fmfchar*}(120,60)
  \fmfleft{tm3,antinu3} \fmflabel{$\tau$}{tm3} \fmflabel{$\nu_{\tau}$}{antinu3}
  \fmfright{g3,f3,fb3} \fmflabel{$\nu_{\ell}$}{fb3} \fmflabel{$\ell$}{f3} \fmflabel{$\gamma $}{g3}
  \fmf{fermion}{tm3,v23,Wi3,antinu3}
  \fmf{dashes,tension=1,label=$W$}{Wi3,Wf3}
  \fmf{photon,tension=0}{v23,g3}
  \fmf{fermion,label=\rotatebox{50}{\vspace{-2mm}\hspace{-3mm}$~$}}{fb3,Wf3}
  \fmf{fermion}{Wf3,f3}
  \fmfdot{Wi3,Wf3,v23}
\end{fmfchar*}
\end{fmffile}\vspace{10mm}
\begin{fmffile}{externalph2}
\begin{fmfchar*}(120,60)
  \fmfleft{tm4,antinu4} \fmflabel{$\tau$}{tm4} \fmflabel{$\nu_{\tau}$}{antinu4}
  \fmfright{g4,f4,fb4} \fmflabel{$\nu_{\ell}$}{fb4} \fmflabel{$\ell$}{f4} \fmflabel{$\gamma $}{g4}
  \fmf{fermion}{tm4,Wi4}
  \fmf{fermion,tension=3}{Wi4,antinu4}
  \fmf{dashes,tension=2.3,label=$W$}{Wi4,Wf4}
  \fmf{fermion,label=\rotatebox{13}{\vspace{-2mm}\hspace{-3mm}$~$}}{fb4,Wf4}
  \fmf{fermion}{Wf4,v24,f4}
  \fmf{photon,tension=0}{v24,g4}
  \fmfdot{Wi4,Wf4,v24}
\end{fmfchar*}
\end{fmffile} \vspace{10mm}
\begin{fmffile}{externalph3}
\begin{fmfchar*}(120,60)
  \fmfleft{tm4,antinu4} \fmflabel{$\tau$}{tm4} \fmflabel{$\nu_{\tau}$}{antinu4}
  \fmfright{g4,f4,fb4} \fmflabel{$\nu_{\ell}$}{fb4} \fmflabel{$\ell$}{f4} \fmflabel{$\gamma $}{g4}
  \fmf{fermion}{tm4,Wi4}
  \fmf{fermion,tension=3}{Wi4,antinu4}
  \fmf{dashes,tension=2.3,label=$W$}{Wi4,vtx}
  \fmf{dashes,tension=2.3,label=$W$,label.dist=-4.5mm}{vtx,Wf4}
  \fmf{fermion,label=\rotatebox{13}{\vspace{-2mm}\hspace{-3mm}$~$}}{fb4,Wf4}
  \fmf{fermion}{Wf4,f4}
  \fmf{photon,tension=0}{vtx,g4}
  \fmfdot{Wi4,Wf4,vtx}
\end{fmfchar*}
\end{fmffile}
\end{center}
\caption{Feynman diagrams for the radiative leptonic decay 
$\tau^- \rightarrow \ell^- \nu_\tau \bar{\nu}_\ell \gamma$ in the SM. 
In the last diagram $\gamma$ is emitted by $W$ boson, the contribution 
of this mechanism is suppressed by a factor of $(m_\tau/m_W)^2\simeq 5\times 10^{-4}$~\cite{w_supp}.}\label{RD} 
\end{figure}

Radiative leptonic decay $\tau^-\to\ell^-\nu_\tau\bar{\nu}_\ell\gamma$ (unless 
specified otherwise, charge-conjugated decays are implied throughout the paper) 
provides an additional promising tool to search for NP. 
Feynman diagrams of this decay in the SM are presented in Fig~\ref{RD}. The presence 
of the radiation exposes the internal structure of $\tau$ decays differently from 
the ordinary leptonic decays. For instance, the measurement of the spectra of outgoing 
lepton and photon allows one to access three more Michel parameters, $\bar{\eta}$, 
$\eta^{\prime \prime}$ and $\xi\kappa$~\cite{oldform,michel_form_arvzov,jibunno}. 
In this note we consider the anomalous four-point $\tau -W-\nu_\tau -\gamma$ scalar 
and tensor couplings. 

\section{Anomalous four-point scalar and tensor interactions} 

\begin{figure}[]
\begin{center}
\begin{fmffile}{anomolus}
\begin{fmfchar*}(120,60)
  \fmfleft{tm3,antinu3} \fmflabel{$\tau$}{tm3} \fmflabel{$\nu_{\tau}$}{antinu3}
  \fmfright{g3,f3,fb3} \fmflabel{$\nu_{\ell}$}{fb3} \fmflabel{$\ell$}{f3} \fmflabel{$\gamma $}{g3}
  \fmf{fermion}{tm3,Wi3,antinu3}
  \fmf{dashes,tension=1,label=$W$,label.dist=-4.5mm}{Wi3,Wf3}
  \fmf{photon,tension=0}{Wi3,g3}
  \fmf{fermion,label=\rotatebox{50}{\vspace{-2mm}\hspace{-3mm}$~$}}{fb3,Wf3}
  \fmf{fermion}{Wf3,f3}
  \fmfdot{Wf3}
  \fmfblob{0.15w}{Wi3}
\end{fmfchar*}
\end{fmffile}\hspace{6mm}
\end{center}
\caption{Feynman diagram for the anomalous four-point $\tau -W-\nu_\tau -\gamma$ interaction.}\label{fptidiagram}
\end{figure}
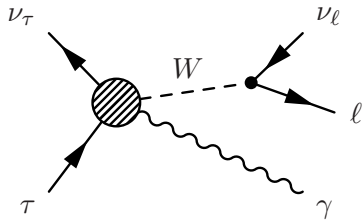

We suggest to consider anomalous four-point $\tau -W-\nu_\tau -\gamma$ 
scalar and tensor interactions (see Fig.~\ref{fptidiagram}), the modified 
Lagrangian of the charged weak interaction of $\tau$ is written as: 
\begin{align}
\mathcal{L} \supset \frac{g}{\sqrt{2}} &W^\mu \Big[ \bar{\psi}(\nu_\tau) \gamma_\mu \frac{1-\gamma^5 }{2} \psi(\tau) -\frac{e\kappa_{\tau}^S}{m_{\tau}} A_{\mu} \bar{\psi}(\nu_\tau) \psi(\tau) 
+\frac{ie\kappa_{\tau}^T}{m_{\tau}} A^{\nu} \bar{\psi}(\nu_\tau) \sigma_{\mu \nu} \psi(\tau)  \Big],   \label{formla_np}
\end{align}
where the first term is the SM Lagrangian, $\kappa_{\tau}^S$ and $\kappa_{\tau}^T$ 
characterize the magnitudes of the scalar and tensor interactions, respectively, 
$A_\mu$ is the electromagnetic field, $\sigma_{\mu \nu} = i[\gamma_{\mu}, \gamma_{\nu}]/2$, 
$e$ is electron charge. 
The introduced terms appear from the gauge invariant dimension-five operators 
$\bar{\psi}(\nu_\tau)D_\mu D^\mu\psi(\tau)$ and $\bar{\psi}(\nu_\tau)\sigma_{\mu \nu} i[D^\mu, D^\nu] \psi(\tau)$, 
where $D_{\mu}$ is the $\mathrm{U}(1)\otimes \mathrm{SU}(2)$ gauge covariant derivative, 
$[D^\mu, D^\nu]\equiv D^\mu D^\nu - D^\nu D^\mu$. 

The total matrix element squared of the $\tau^-\rightarrow \ell^- \nu_\tau \bar{\nu}_\ell \gamma$ 
decay in the presence of the anomalous amplitude $\mathcal{M}_{N}~({N}=S,~T)$ is written as: 
\begin{align}
 |\mathcal{M}_\mathrm{tot}|^2 &= |\mathcal{M}_\mathrm{SM}+\mathcal{M}_{N}|^2 \nonumber \\
 &= |\mathcal{M}_\mathrm{SM}|^2 + 2\Re\{\mathcal{M}_\mathrm{SM}\mathcal{M}_{N}^* \} + |\mathcal{M}_{N}|^2. \nonumber
\end{align}
The contribution to the differential decay width from the $2\Re\{\mathcal{M}_\mathrm{SM}\mathcal{M}_{N}^* \} + |\mathcal{M}_{N}|^2$ is given by: 
\begin{align}
\frac{\mydif \Gamma_N(\tau^- \rightarrow \ell^- \nu_\tau \bar{\nu}_\ell \gamma )}{ \mydif x \mydif y \mydif\Omega_{\ell}^* \mydif\Omega_{\gamma}^*  }= \frac{4 m_{\tau}^5 G_{F}^2  \alpha }{3 (4\pi)^6 }
\frac{x\beta_{\ell}^*}{z} \left[ \kappa^N_\tau F_1^N(x,y,z) + (\kappa^N_\tau)^2 F_2^N(x,y,z) \right], \label{dff}
\end{align}
where $G_F = 1.1663787(6)\times 10^{-5}$~GeV$^{-2}$\cite{PDG_paper} is the Fermi constant and $\alpha = 1/137.035999139(31)$\cite{PDG_paper} is the fine-structure constant, $p_\ell = (E_\ell,\vec{p}_\ell)$ and $p_\gamma = (E_\gamma ,\vec{p}_\gamma)$ are 
four-momenta of the outgoing charged lepton and photon, respectively; $\beta_{\ell}=|\vec{p}_\ell|/E_\ell$, $\Omega_\ell$ and 
$\Omega_\gamma$ are the solid angles of the final charged lepton and photon, respectively;  
$x=2E_{\ell}^*/m_\tau$, $y=2E_{\gamma}^*/m_\tau$, $z=2 (p_{\gamma}\cdot p_{\ell})/m_\tau^2 = xy(1-\beta_{\ell}^* \cos\theta_{\ell \gamma}^*)/2$, $\cos\theta_{\ell \gamma}=(\vec{p}_\ell\cdot \vec{p}_\gamma)/(|\vec{p}_\ell|\cdot|\vec{p}_\gamma|)$ (asterisks indicate parameters measured in the $\tau$ rest frame) and $\lambda=m_\ell/m_\tau$; $F_{1}^N$ and $F_{2}^N$ are the form factors 
(see Appendix for the explicit formulae). 
Integrating the differential decay width numerically, we obtain: 
\begin{align}
&\Gamma(\tau^- \rightarrow \ell^- \nu_\tau \bar{\nu}_\ell \gamma)_{E_{\gamma}^*>10~\mathrm{MeV}}= \Gamma^\mathrm{SM}_{E_{\gamma}^*>10~\mathrm{MeV}} \left[1+c^N_\ell\kappa^N_\tau +d^N_\ell(\kappa^N_\tau)^2  \right], \nonumber
\end{align}
where the coefficients $c_{\ell}^{N}$ and $d_{\ell}^{N}$ $(N=S$ or $T)$ are:
$c_{e}^S = (-2.06\pm 0.01)\times 10^{-3}$, $c_{\mu}^S = (-8.46\pm 0.03)\times 10^{-3}$, 
$c_{e}^T = (-6.18\pm 0.02)\times 10^{-3}$, $c_{\mu}^T = (-3.19 \pm 0.006)\times 10^{-2}$,
$d_{e}^S =  (3.77\pm 0.02)\times 10^{-4}$, $d_{\mu}^S =  (1.85 \pm 0.006)\times 10^{-3}$, 
$d_{e}^T = (4.51\pm 0.03)\times 10^{-3}$ and $d_{\mu}^T = (2.21\pm 0.006)\times 10^{-2}$,
where the error is statistical uncertainty of numerical integration. 
In this calculation, we use
 the photon energy threshold of $10~\mathrm{MeV}$ in the $\tau$ rest frame.
 Taking into account the SM prediction 
$\mathcal{B}^{\rm Exc}(\tau^- \rightarrow \mu^- \nu_\tau \bar{\nu}_\mu\gamma)=(3.572\pm 0.007)\times 10^{-3}$~\cite{radiativedecay_theory}
 and the PDG average $\mathcal{B}(\tau^- \rightarrow \mu^- \nu \bar{\nu} \gamma)=(3.68\pm 0.10)\times 10^{-3}$~\cite{PDG_paper},
we get the constraint at 95\% CL: $-0.025<(c^N_{\mu}\kappa^N_\tau +d^N_{\mu} (\kappa^N_\tau)^2)<0.085$, or: 
\begin{align}
-4.9 < {\kappa_{\tau}^S} &<9.4~~~~( 95\%~{\rm CL}),\\
-1.4 < {\kappa_{\tau}^T} &<2.8~~~~( 95\%~{\rm CL}).
\end{align}
Moreover, the four-point $\tau -W-\nu_\tau -\gamma$ scalar and 
tensor interactions can be searched for at the electron-positron 
colliders, in the process of the $e^+ e^-$ annihilation 
$e^+e^-\to\gamma^*\to\tau^-\bar{\nu}_\tau (W^+\to\ell^+\nu_\ell,{\rm hadron(s)}^+)$. 
This anomalous coupling is responsible for the production of the 
single $\tau$ lepton below $\tau^+\tau^-$ production threshold. 
Such a mechanism can be searched for at the low energy $e^+ e^-$ colliders like 
Beijing Electron-Positron Collider (BEPC), Cornell Electron Storage Ring (CESR) 
and VEPP-2000~\cite{vepp1,vepp2}, as well as at the B-factories, 
Belle~\cite{bel}/KEKB~\cite{kekb} and $BABAR$~\cite{Aubert:2001tu}/PEP-II, in the processes 
of $e^+ e^-$ annihilation with the initial state radiation~\cite{Bevan:2014iga}. 
For example, at VEPP-2000 in the center-of-mass energy range from about $1.8$~GeV 
up to $2.0$~GeV (near the production threshold of single tau), $\tau^-$ lepton is 
produced almost at rest, accompanied with $e^+\nu_e \bar{\nu}_\tau$, $\mu^+\nu_\mu\bar{\nu}_\tau$ or $\pi^+\bar{\nu}_\tau$. 
As a result, besides $e-\mu$ events, a clear signature 
of the production of single $\tau^-$ will be the monochromatic $\pi^-$ or $\rho^-$ 
(from $\tau^-\to\pi^-\nu_\tau$ or $\tau^-\to\rho^-\nu_\tau$ decays). 

\section{Summary}

The precision measurement of the properties of leptonic decays of $\tau$ lepton 
offers unique opportunity to search for the physics beyond the Standard Model.
The radiative leptonic decay $\tau^- \rightarrow \ell^- \nu_\tau \bar{\nu}_\ell \gamma$ 
provides an additional tool to probe the internal structure of the weak interaction.
The anomalous four-point $\tau -W-\nu_\tau -\gamma$ scalar and tensor interactions 
are simple extensions of the Standard Model, which affect the spectra of the daughter 
particles in the radiative leptonic decays of tau. 
We calculated the corresponding differential and the total decay widths. 
The world average value of the branching ratio of the $\tau^- \rightarrow \mu^- \nu_\tau \bar{\nu}_\mu \gamma$ 
decay and its SM prediction constrain the magnitudes of the scalar and tensor couplings 
to be $-4.9 < \kappa_{\tau}^S <9.4$ and $-1.4 < \kappa_{\tau}^T<2.8$ (at 95\% CL), respectively. 
In this note, we extract the constraints on the $\kappa_\tau^S$ and $\kappa_\tau^T$ coupling 
constants from the total widths, but the more sensitive method is to fit the full 
differential decay width of the radiative leptonic decay of tau. 

\section*{Acknowledgments}

We are grateful to Andrey Pomeransky (Budker Institute of Nuclear Physics) for the fruitful discussions.

\appendix
\section{Form factors}
\begin{align}
F_{1}^S &= -z \big\{ (1+\lambda^2 -x-y+z)(z-3x) \nonumber \\
   &\hspace{3cm} + ( y-z)(x-z-2\lambda^2 )  \big\} \nonumber \\
 & \hspace{3cm} + 3y(z-2\lambda^2)(1+\lambda^2-x-y+z), \nonumber   \\
F_{2}^S &= 2xz(1+\lambda^2-x-y+z)+2z(2-y-x)(x-z-\lambda^2),\nonumber 
\end{align}
\begin{align}
F_{1}^T &= z(-3x+x^2+13xy-9y+9y^2) +z^2(-7x-17y+7)  \nonumber \\
 &+ 6z^3  \lambda^2  \{ -18y+18xy+18y^2 + z(8-3x-37y)  \nonumber \\
 &+9z^2  \} -18\lambda^4 y, \nonumber \\ 
F_{2}^T &= 26xz(1+\lambda^2-x-y+z)+2z(2-y-x)(x-z-\lambda^2).\nonumber 
\end{align}

\end{document}